\documentclass[onecolumn,useAMS,usenatbib]{mn2e}
\usepackage{amssymb,graphicx,subfigure,color}

\newcommand{\cs}{c_{\rm s}}

\newcommand{\Ledd}{L_{\rm E}}

\newcommand{\rs}{R_{\rm S}}

\newcommand{\Teff}{T_{\rm eff}}

\newcommand{\cm}{\,{\rm cm}}
\newcommand{\g}{\,{\rm g}}
\newcommand{\K}{\,{\rm K}}

\newcommand{\kms}{\,{\rm km~s^{-1}}}
\newcommand{\kpc}{\,{\rm kpc}}
\newcommand{\pc}{\,{\rm pc}}

\newcommand{\prad}{p_{\rm{rad}}}
\newcommand{\pgas}{p_{\rm{gas}}}
\newcommand{\Qmin}{Q_{\rm{min}}}
\newcommand{\rmin}{r_{\rm{min}}}
\newcommand{\rmax}{r_{\rm{max}}}
\newcommand{\um}{\,\mu\rm{m}}
\newcommand{\Mdot}{\dot{M}}
\newcommand{\lum}{l_{\rm{E}}}

\title{Spectral energy distributions of selfgravitating QSO discs}
\author[E. Sirko and J. Goodman]
       {Edwin Sirko and Jeremy Goodman\\
    Princeton University Observatory, Princeton, NJ 08544, USA}
\date{Received ??}

\pagerange{\pageref{firstpage}--\pageref{lastpage}}
\pubyear{2002}

\begin{document}

\maketitle

\label{firstpage}

\begin{abstract}
We calculate spectral energy distributions (SEDs) of steady accretion
discs at high accretion rates, as appropriate for bright QSOs, under
the assumption that the outer parts are heated sufficiently to maintain
marginal gravitational stability, presumably by massive stars formed within
the disc.  The SED is independent of the nature of these auxiliary
sources if their inputs are completely
thermalized.  Standard assumptions are made for angular momentum
transport, with an alpha parameter less than unity.  With these
prescriptions, the luminosity of the disc is sensitive to its opacity,
in contrast to standard discs powered
by release of orbital energy alone.  Compared to the latter, our discs
have a broader SED, with a second peak in the
near-infrared that is energetically comparable to the blue bump.
The energy in the second peak increases with the outer radius of the disc,
provided that the accretion rate is constant with radius.
By comparing our computed SEDs with observed ones,
we limit the outer radius of the
disc to be less than $10^5$ Schwarzschild radii ($R_{\rm S}$),
or about one parsec, in a typical QSO.
We also discuss some properties of our minimum-$Q$
discs in the regions where auxiliary heating is dominant
($10^3-10^5 R_{\rm S}$).

\end{abstract}


\section{Introduction}

The standard
theoretical paradigm for the central engine in quasars
and their radio-quiet kin (QSOs) is a viscous accretion
disc surrounding a massive black hole.  
Direct evidence for accretion discs, such as
double-peaked emission lines \citep{Eracleous_Halpern94},
is important to seek but hard to come by.
Perhaps the best reasons
for belief in this paradigm are basic considerations of energy
and angular momentum.  Such discs are the most plausible
astrophysical mechanism for converting rest mass to
radiation with high efficiency, and if the relics of QSOs reside
in galactic nuclei, then efficiencies $\gtrsim 10\%$ are required
\citep{Soltan82,Chokshi_Turner92,Yu_Tremaine02}.
On the other hand, the maximum specific angular momentum
of a Kerr black hole, $GM/c\approx 1.4 M_8\kms$, is far less than
that of most mass in galaxies. ($M_8=M/10^8 M_\odot$, where
$M$ is the mass of the black hole.)
Accreting gas must be separated from its angular momentum,
and a viscous disc is the most natural mechanism.

Most of the binding energy of the disc is in its inner parts, so that
the outer radius of a standard accretion disc is almost irrelevant to
its bolometric luminosity.  In contrast, most of the angular momentum
is in the outer regions, so that the outer radius of the disc is
sensitive to the initial angular momentum of the gas that feeds it.
In a previous paper \citep[][henceforth Paper I]{Goodman02}, it was
argued that luminosity and angular momentum are coupled by the
requirement that QSO discs be stable against their own self-gravity.

It is well known that gravitational stability is problematic
in the outer parts of QSO discs \citep{Shlosman_Begelman87}.
The threat to the standard paradigm is that a strongly selfgravitating
disc is likely to fragment completely into stars, leaving
insufficient gas to fuel the QSO.  As discussed in Paper I, viable
solutions to this difficulty fall into several categories:
\begin{enumerate}
\item enhanced angular momentum transport, not necessarily by viscous
processes but at rates corresponding to
a viscosity parameter $\alpha\gg 1$;
\item auxiliary heating in excess of what is provided
by dissipating orbital energy, so as to reduce the gas density and
selfgravity of the disc;
\item replacement of the outer disc with a very
dense star cluster, whose collisional debris supply a disc of
small radius and negligible selfgravity;
\item relatively low initial angular momentum for the gas, which therefore
circularizes at small enough radius so as to avoid selfgravity.
\end{enumerate}
A number of options in the first
category were briefly considered in Paper I,
including accretion driven by bars or global spiral waves
\citep[e.g.][]{Shlosman_Begelman89} or by magnetized winds
\citep[e.g.][]{Blandford_Payne82}.  But it was argued that none of these
options is likely to achieve a supersonic accretion speed, and hence
that discs cannot be stabilized much beyond one parsec by any of these
mechanisms alone.  The third category was tentatively rejected on the
grounds that remnants of such star clusters are not observed
in present day galactic nuclei.  Recently, \cite{Pariev_etal02} have
proposed discs whose thickness is supported primarily by magnetic
pressure.  If this is possible, it would imply lower disc densities
on average, but the field might well squeeze the gas into dense clumps
and thereby actually exacerbate selfgravity.  At any rate,
\cite{Pariev_etal02} do not apply their model to the cool outer regions
beyond $10^3\rs$.

The present paper will focus on some implications of the second
category of solutions, which appear to be a natural compromise between
a purely gaseous disc and a star cluster.  It is reasonable to suppose
that part of the gas fragments into stars, and that the energy
released by nuclear fusion and other stellar processes (supernovae,
stellar-mass black holes) may sufficiently heat the rest of the gas so
as to prevent complete fragmentation
\citep{Collin_Zahn99a,Collin_Zahn99b}.  There is a great deal of
evidence that such a feedback cycle operates in the
discs of spiral galaxies on kiloparsec scales.

Observed SEDs of typical quasars differ markedly from classical
theoretical predictions in which the disc is assumed to be
geometrically thin, optically thick, steady, and heated solely by
viscous dissipation \citep{Pringle81}.  To a first approximation, the
typical SED is flat in a $\lambda F_\lambda$ plot over many decades in
wavelength \citep{Elvis_etal94}.  Relative to the classical
predictions, there is excess emission at both X-ray and infrared
wavelengths.  The former is conventionally ascribed to comptonization
in a hot corona at small radii \citep{Shapiro_Lightman_Eardley76}, and
the latter to passive reprocessing in warped or flared outer parts of
the disc \citep{Sanders_etal89}.

We suppose that the infrared excess may be due to the energy inputs
required to stabilize the outer disc against its own selfgravity.  We
obtain a lower bound on the auxiliary heating needed to stabilize the
disc for given values of the macroscopic parameters: namely, the black
hole mass ($M$), accretion rate ($\dot M$), and disc outer radius
($r_{\max}$).  As shown in Paper I, and confirmed here with more
realistic opacities, the inputs required for gravitational
stability increase with the outer radius of the disc, $r_{\rm max}$.
Paper I argued for an upper limit to $r_{\rm max}$ based on the
energy available from plausible sources, such as fusion or accretion
onto stellar-mass black holes.  In this paper, we find limits to
$r_{\rm max}$ from the SED.

It is not obvious that the auxiliary inputs should be
completely thermalized, but if we assume this, then the SED can be
predicted.  The observed infrared emission of typical QSOs may be due largely
to reprocessing of light emitted from the inner parts of the disc,
as conventionally supposed.  But by attributing all of the infrared
light to the auxiliary energy inputs, and insisting that these
inputs be sufficient for gravitational stability, we obtain bounds for
$r_{\rm max}$.  These bounds depend upon other parameters, especially
the mass of the black hole, the accretion rate, and 
the viscosity parameter $\alpha$.  We explore these dependencies.
The meaning of $r_{\max}$ constrained by this method is the
radius within which $\dot M$ is sensibly constant; obviously the disc
can be extended indefinitely if the accretion rate and mass at large
radii are sufficiently small.  Hence our limits on $r_{\max}$ are best
translated into upper limits on the initial angular momentum of the
gas that is accreted.

The outline of our paper is as follows.  \S 2 lays out the physical
assumptions and governing equations for our disc models.  These are
the same as for the classical steady thin disc, with the
one important exception that wherever the classical model would
be gravitationally unstable, we invoke just enough auxiliary heating
to stabilize it.  A discussion of opacities becomes critically
important, because the requirement of marginal stability fixes the density
and temperature at the midplane (for given $\dot M$ and $\alpha$); the flux
escaping from the disc, and hence the amount of auxiliary heating needed,
then depend upon the optical depth.
The computed radial structure of the disc is presented
in \S3 for parameters representative of bright quasars.
Since our assumptions are no different from the classical ones at
small radii, we emphasize the
properties of our discs in the marginally selfgravitating
region $r\gtrsim 10^3\rs\sim 10^{-2}\pc$, and we compare our SEDs with
those presented by \cite{Elvis_etal94} to obtain upper limits
on the outer radius of the disc and the initial angular momentum of
the gas.  In the final section, we summarize our conclusions, issue
the necessary caveats, and discuss directions for future research.

\section{The selfgravitating accretion disc model}\label{s:model}

We start by considering the standard steady $\alpha$ disc.  In
this model, gas is accreted onto a central black hole of mass $M
\sim M_8$ in a steady, Keplerian, geometrically thin 
accretion disc at a constant rate $\dot{M}$. Accretion is driven
by viscous mechanisms, but the exact mechanism (probably
magnetorotationally-driven turbulence) is unimportant here.
Newtonian equations will be used.  These are not accurate near the
inner edge of the disc, but the effects of selfgravity are at
large radii where relativistic corrections are unimportant.

We assume that each annulus of
the disc radiates as a blackbody with temperature $\Teff(r)$.
When accretion is the only source of thermal energy,
\begin{equation}\label{e:Teff-Mdot}
\sigma\Teff^4 = \frac{3}{8\pi}\dot{M}'\Omega^2,
\end{equation}
where $\Omega = (GM/r^3)^{-1/2}$, and we have defined for
notational simplicity $\dot{M}' = \dot{M}(1-\sqrt{\rmin/r})$ where
$\rmin$ is the inner radius of the accretion disc. A zero-torque
boundary condition has been assumed. In a relativistic thin-disc
analysis, $\rmin$ would be the radius of the marginally stable
orbit.  In our newtonian framework, we set $\rmin=\rs/4\epsilon$,
where $\rs=2GM/c^2$ is the Schwarzschild radius and $\epsilon\sim
0.1$ is the assumed radiative efficiency of accretion.

An $\alpha$ prescription for the viscosity is adopted,
$\nu=\alpha\cs h\beta^b$, where $\cs=\sqrt{p/\rho}$ is the isothermal
sound speed at the disc midplane, $h=\cs/\Omega$ the half thickness,
and $\beta = \pgas/p$ is the ratio of gas pressure to total pressure at
the midplane.  The viscosity is related to the accretion rate by
\begin{equation}\label{e:Sigma-alpha}
\beta^b \cs^2 \Sigma = \frac{\dot{M}'
\Omega}{3 \pi \alpha},
\end{equation}
where $\Sigma = \int_{-\infty}^{+\infty} \rho \, dz$ is the
surface density at a given radius. The parameter $b$ is a switch
that can be either $1$, so that viscosity
is proportional to gas pressure, or $0$, so that viscosity is
proportional to total pressure.  Magnetic pressure is assumed to
contribute negligibly to the thickness of the disc, even if magnetic
stresses dominate angular momentum transport.

Because it will turn out that the optically thick assumption fails at
large radii, we now turn to a study of the relationship between
the midplane temperature $T$ and $\Teff$.  We assume radiative
energy transport in the vertical ($z$) direction and write
$\tau=\int_0^\infty \kappa\rho dz$ for the optical depth at the midplane.
In the optically thick limit $\tau\gg 1$, the diffusive approximation
is applicable so that $T^4\sim\tau\Teff^4$, with a numerical factor
of order unity that depends upon the vertical variation
of opacity $\kappa$ and the vertical distribution of viscous dissipation.
The latter is not yet predictable from theory, despite major advances
in understanding magnetohydrodynamic disc turbulence.  If one assumes
constant dissipation per unit optical depth, then
\begin{displaymath}
T^4\approx\left(\frac{3}{8}\tau+\frac{1}{2}\right)\Teff^4,
\end{displaymath}
where the ``$+1/2$'' follows from the Eddington approximation to
the boundary condition at the photosphere.  In the opposite limit
of low optical depth, the radiative flux from
one side of the disc scales as $\sigma\Teff^4\sim\tau\sigma T^4$.
Again a numerical factor occurs.  One should distinguish
between absorption and scattering opacity, but at the large accretion
rates of interest to us, optically thin conditions tend to
occur where the opacity is primarily absorptive.  For a vertically
isothermal disc, $\sigma\Teff^4=4\tau\sigma T^4$ when $\tau\ll 1$.
Simple interpolation between the optically thin and optically
thick limits yields eq.~(\ref{e:T}) below, which satisfies the
physical requirement that $\Teff\le T$ for any $\tau$ (in fact,
the minimum of $T^4/\Teff^4$ is $1.112$ and occurs when 
$\tau = \sqrt{2/3}$).

We represent the effective radiation pressure by eq.~(\ref{e:prad}).
In the optically thick case, in view of eq.~(\ref{e:T}),
this reduces to the usual $\prad = aT^4/3$
relation, but in the optically thin case,
$\prad \to 2\tau^2\sigma T^4/c$: one power of $\tau$ represents
the inefficiency with which photons are radiated, and the second is the
fraction of these photons that transfer their momentum to the gas, thereby
helping to support the thickness of the disc.  The numerical factor is
correct for an isothermal slab of midplane  optical depth $\tau\ll1$.

\begin{figure}
\includegraphics[width=\textwidth]{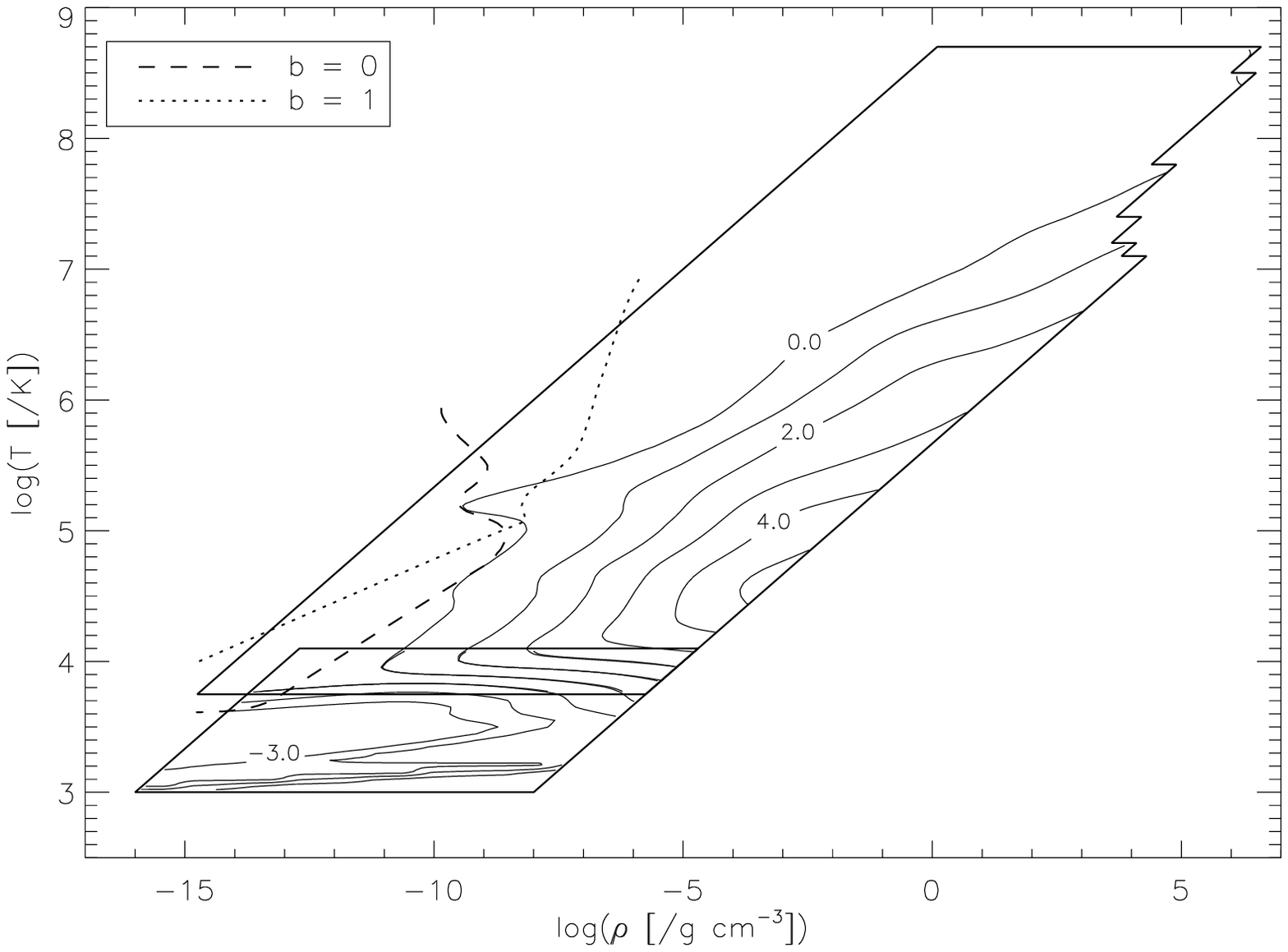}
\caption{Contour plot of opacity as a function of density and
temperature, as provided by \protect{\citet{iglesias_rogers_1996}} (high-$T$
region) and \protect{\citet{alexander_ferguson_1994}} (low-$T$ region) for
$X=.70$, $Z=.03$.  Contours represent integer values of the logarithm
of the opacity ($\rm{cm^2\,g^{-1}}$).  The overlap region 
$3.75 < \log{T} < 4.1$ is plotted with contours from both sources; their 
agreement is illustrated.
The dashed ($b=0$) and dotted ($b=1$) lines show
trajectories of $(\rho,~T)$ for typical examples of our disc models
(Figure~\protect{\ref{f:radial_parms}} for $b=0$), up to $\rmax = 10^5 \rs$.}
\label{f:opacity}
\end{figure}

We also must know the opacity in the disc
(equation~\ref{e:kappa}).  We use the opacity tables of
\citet{iglesias_rogers_1996} for high temperatures and those of
\citet{alexander_ferguson_1994} for low temperatures, 
with $X=.70$, $Z=.03$. Figure~\ref{f:opacity} illustrates the opacity
as a function of density and temperature.  The two opacity tables
overlap in the region $3.75 < \log{T} < 4.1$, and 
Figure~\ref{f:opacity} demonstrates
their agreement. Additionally, the loci
of two disc models are projected onto 
the plot.  Extrapolation of the opacity tables in the low density
regime is necessary.  This is done by simply assigning $\kappa$  
its value at the low-density edge of the opacity table
(using the same value of $T$).
This extrapolation is presumably correct for 
high temperatures ($T \gtrsim 10^4 \K$) where the opacity is dominated by
electron scattering.  At lower temperatures, the extrapolation
is less certain, but we feel it is reasonable as a first approximation.
Also, for $\log{T} < 3$, $\kappa$ is
assigned its value at the low-temperature edge of the opacity table,
which is approximately $10^{0.76} \cm^2\g^{-1}$.

In summary, our governing
equations for the disc are as follows (we take $m=.62~m_H$ as the mean
molecular mass):
\begin{eqnarray}
\label{e:Teff}\sigma \Teff^4 &=& \frac{3}{8\pi} \dot{M}' \Omega^2 \\
\label{e:T}
T^4 &=& \left(\frac{3}{8}\tau + \frac{1}{2} + \frac{1}{4\tau}\right) \Teff^4 \\
\label{e:tau}\tau &=& \frac{\kappa \Sigma}{2} \\
\label{e:alpha}\beta^b \cs^2 \Sigma &=& \frac{\dot{M}' \Omega}{3\pi\alpha} \\
\label{e:prad}\prad &=& \frac{\tau \sigma}{2c} \Teff^4 \\
\label{e:pgas}\pgas &=& \frac{\rho k T}{m} \\
\label{e:beta}\beta &=& \frac{\pgas}{\pgas + \prad} \\
\label{e:Sigma}\Sigma &=& 2 \rho h \\
\label{e:h}h &=& \frac{\cs}{\Omega} \\
\label{e:cs}\cs^2 &=& \frac{\pgas + \prad}{\rho} \\
\label{e:kappa}\kappa &=& \kappa(\rho, T)
\end{eqnarray}

For an accretion disc with given parameters $M$, $\dot{M}$, $\alpha$,
and $b$, we can solve these eleven equations for the eleven unknowns:
$\Teff$, $T$, $\tau$, $\Sigma$,
$\beta$, $\cs$, $\prad$, $\pgas$, $\rho$, $h$, and $\kappa$ at
every radius $r$, which is related simply to $\Omega$.

The outer parts of the accretion disc will be prone to selfgravity if
Toomre's stability parameter, 
\begin{equation}\label{e:toomre}
 Q \equiv
\frac{\cs\Omega}{\pi G \Sigma} \approx \frac{\Omega^2}{2 \pi G \rho}
\end{equation}
is $\lesssim 1$.  In our model, we assume there is some feedback
mechanism, perhaps star formation, that supplies just enough
additional heat to the disc as to prevent $Q$ falling below a minimum
value $\Qmin\approx 1$.  The energy source for these auxiliary inputs
must be something other than the orbital energy of the gas, for
example, nuclear fusion in stars.  The quantity of these inputs is
fixed by the properties of the disc: in particular the opacities,
since the more easily radiation escapes, the more the disc must be
heated to maintain stability.  The nature of the auxiliary sources is
not important to the spectral energy distribution as long as their
inputs are completely thermalized.

When $Q=\Qmin$, eq.~(\ref{e:Teff}), which assumes the disc to be heated
by release of orbital energy only, is no longer appropriate;
we replace it with
\begin{equation}
\label{e:rho-Qmin} \rho = \frac{\Omega^2}{2 \pi G \Qmin}
\end{equation}
and $\Teff$ must be obtained from equation~\ref{e:T}.

Once $\Teff$ is known as a function of $r$, we compute a theoretical
spectral energy distribution for the disc as a weighted sum of Planck
functions:
\begin{equation}\label{e:planck}
L_\lambda = \frac{2 h c^2}{\lambda^5} \int_{\rmin}^{\rmax} 2 \pi r \,dr
    \frac{1}{\exp{[h c/\lambda k \Teff(r)]} - 1}
\end{equation}
This formula does not allow for differences between effective and
color temperature.  Such differences are most likely
in those parts of the disc whose emission is dominated by passive reprocessing
\citep{Sanders_etal89} or are optically thin (to absorption).  In
fact we find that much of the auxiliary heat input occurs in regions
that are marginally optically thick, since these regions cool most
efficiently.
We feel that eq.~(\ref{e:planck}) is adequate for a
first investigation:  it may somewhat overestimate the wavelength
of the infrared peak associated with the auxiliary heating, but the
area under the peak should be roughly correct.

\section{Results}\label{s:results}

We define $\lum \equiv L_0/\Ledd$, where $L_0 \equiv \epsilon \Mdot c^2$ is
the luminosity corresponding to a nonselfgravitating (i.e., with
no auxiliary heating) disc with accretion
rate $\Mdot$, and $\Ledd$ is the Eddington luminosity.  
In our models we always set $\epsilon=.1$ and $\Qmin=1$.
 
\begin{figure}
\includegraphics[width=\textwidth]{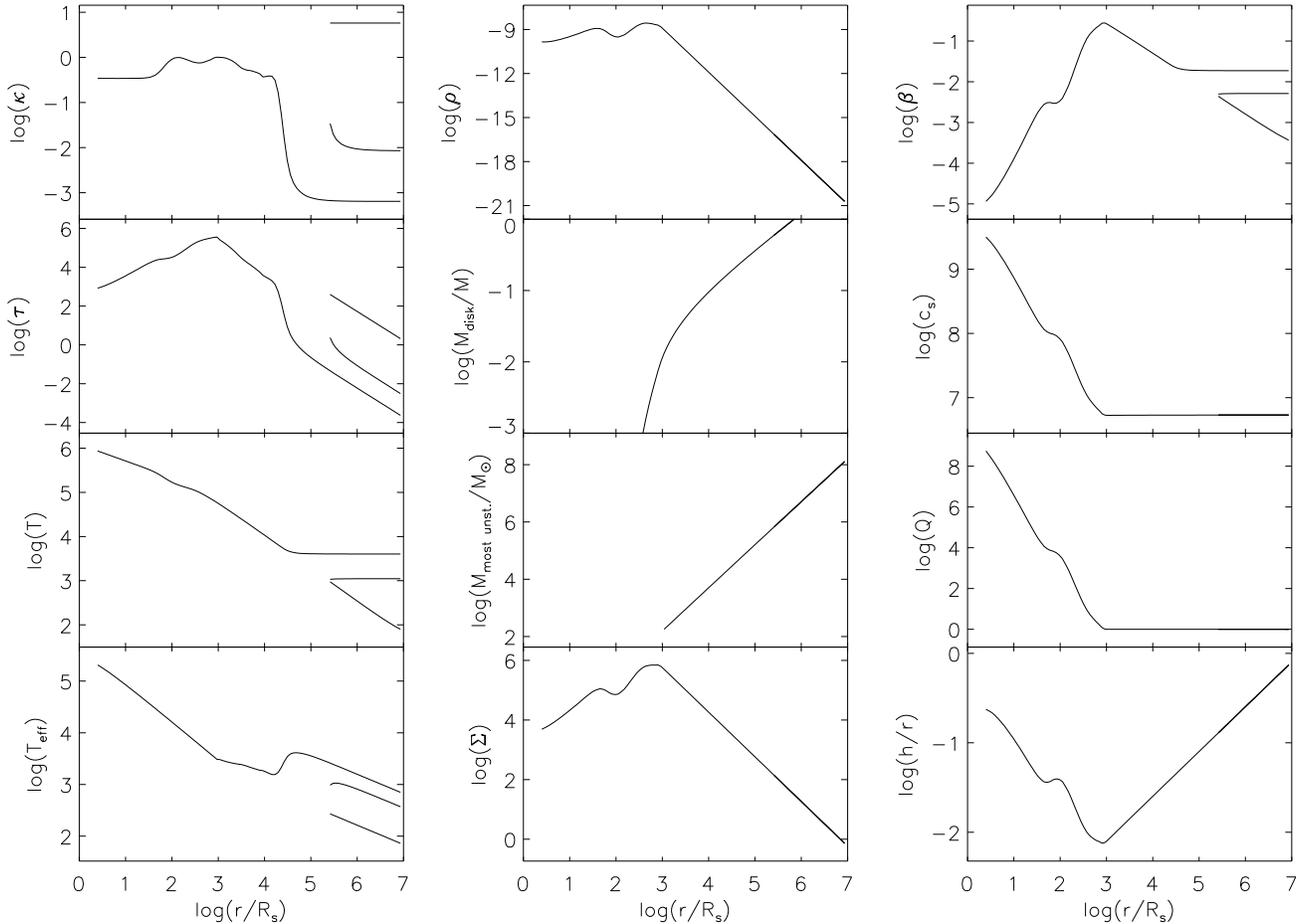}
\caption{Dependence of an assortment of parameters on radius for our
$b=0$ (\emph{i.e.}, viscosity $\propto$ total pressure) canonical case
of $M_8 = 1$, $\lum = .5$, $\alpha = .01$, and $\Qmin = 1$.  All units are in
cgs except for those represented as reduced variables.  The 
integrated disc mass $\int_{\rmin}^r 2\pi r'\Sigma dr'$ is in units of
the central black hole mass $M$, but the most unstable mass 
$\cs^4/G^2\Sigma$ is in units of solar masses.
The SED for this model is the vertically-hatched region in
Figure~\protect{\ref{f:sed_rmax_b0}}. }
\label{f:radial_parms}
\end{figure}

The radial structure of a typical accretion disc ($M_8=1$, $\lum=.5$,
$\alpha=.01$) is presented in Figure~\ref{f:radial_parms} for
$b=0$.  The outer radius of $\rmax=10^7 
\rs$ will be shown to be unrealistic, but this model illustrates 
several key regimes in the disc. 

The inner region ($2.5 \lesssim r/\rs \lesssim 10^2$) is radiation
dominated ($\beta \ll 1$) and the temperature is high enough to ensure
that the opacity is dominated by electron scattering, hence constant.
In view of equation~\ref{e:Teff-Mdot}, $\Teff \propto r^{-3/4}$.  When
$Q$ would otherwise drop below $1$ (here, at $r \approx 10^3\rs$), we
hold $Q$ constant, which ensures that $\rho \propto r^{-3}$.  In this
auxiliary heating region, if $\kappa = \rm{constant}$, then $\Teff
\propto r^{-3/8}$.  After the disc becomes optically thin ($\tau < 1$) we
find multiple solutions.  The one with the least optical depth
corresponds to the continuous line in all panels; it reaches constant
$T$ and $\kappa$.  There is an optically thick solution with $T \propto
r^{-3/4}$.  Since the region in which we find multiple solutions is so
dependent on the extrapolation of opacity into low-density and
low-temperature regimes, the interpretation must be approached with
caution.  However, the presence of multiple solutions raises the
possibility of thermal instability and time-dependence \citep[see, for
example,][]{siemiginowska_etal_1996}.  Note that in our solutions,
$\Mdot$ is constrained to be constant throughout the disc.
Several of the panels in Fig.~\ref{f:radial_parms} contain only
one curve in the region of multiple solutions ($r\gtrsim 10^5\pc$):
this is because $\rho$ and $\cs$ are uniquely determined by
$r$, $\dot M$, $\alpha$, and $Q$ \emph{via} eqs.~(\ref{e:rho-Qmin})
and (\ref{e:alpha}) when $b=0$.

For $b=1$, the effective temperature
has the same $\Teff \propto r^{-3/4}$ behavior
in the inner part of the disc, but differs from the $b=0$ case in the 
auxiliary heating region.  For $\kappa \propto \rm{constant}$ in this
region, in the 
optically thick solution, $T\propto r^{-1/2}~\&~\Teff \propto r^{-1/4}$,
and in the optically thin case,
$T\propto\mbox{const.}~\&~\Teff \propto r^{-3/8}$.

\begin{figure}
\subfigure[b=0]{
\label{f:sed_rmax_b0}
\begin{minipage}{.5\textwidth}
\includegraphics[width=\textwidth]{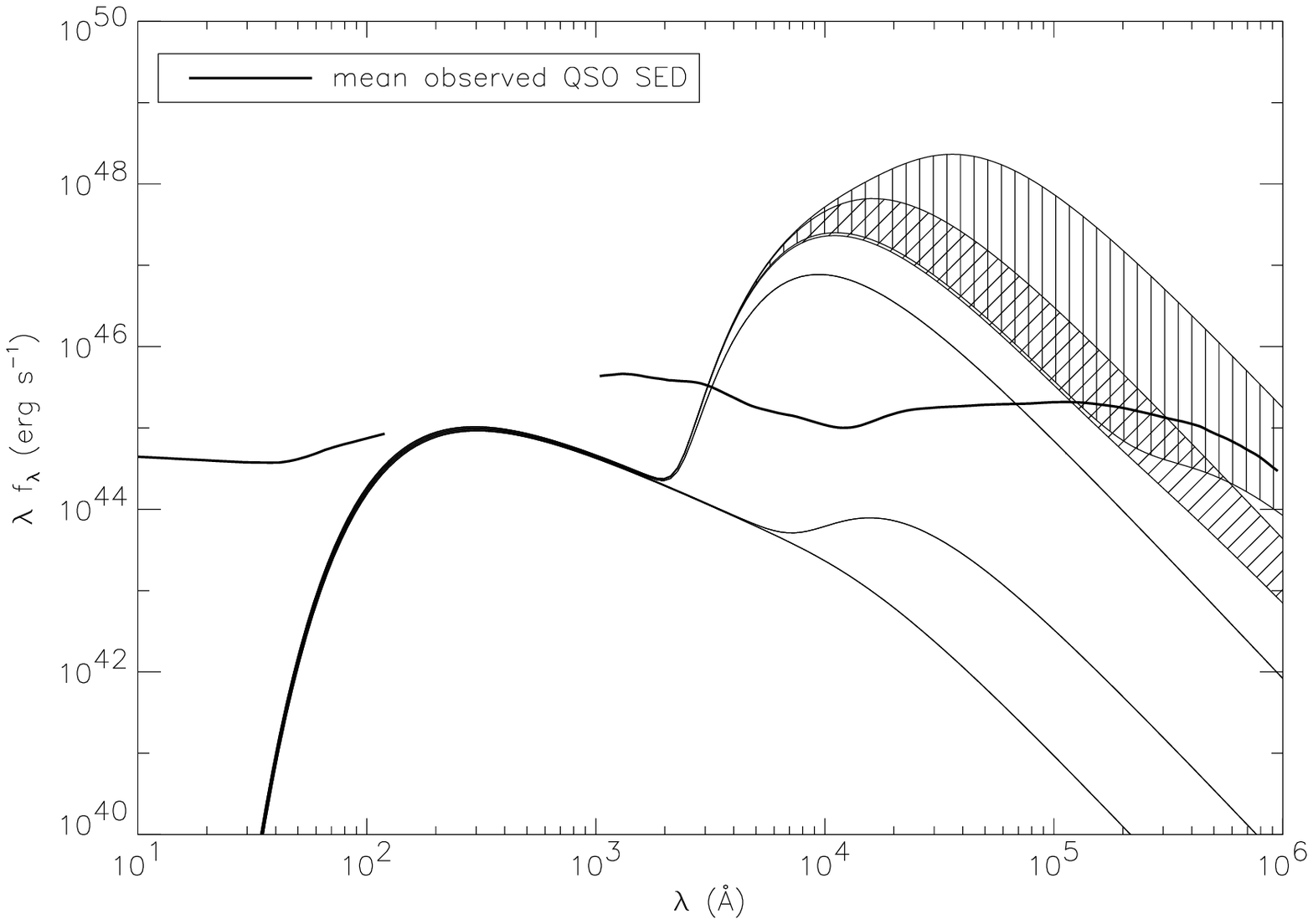}
\end{minipage}}%
\subfigure[b=1]{
\label{f:sed_rmax_b1}
\begin{minipage}{.5\textwidth}
\includegraphics[width=\textwidth]{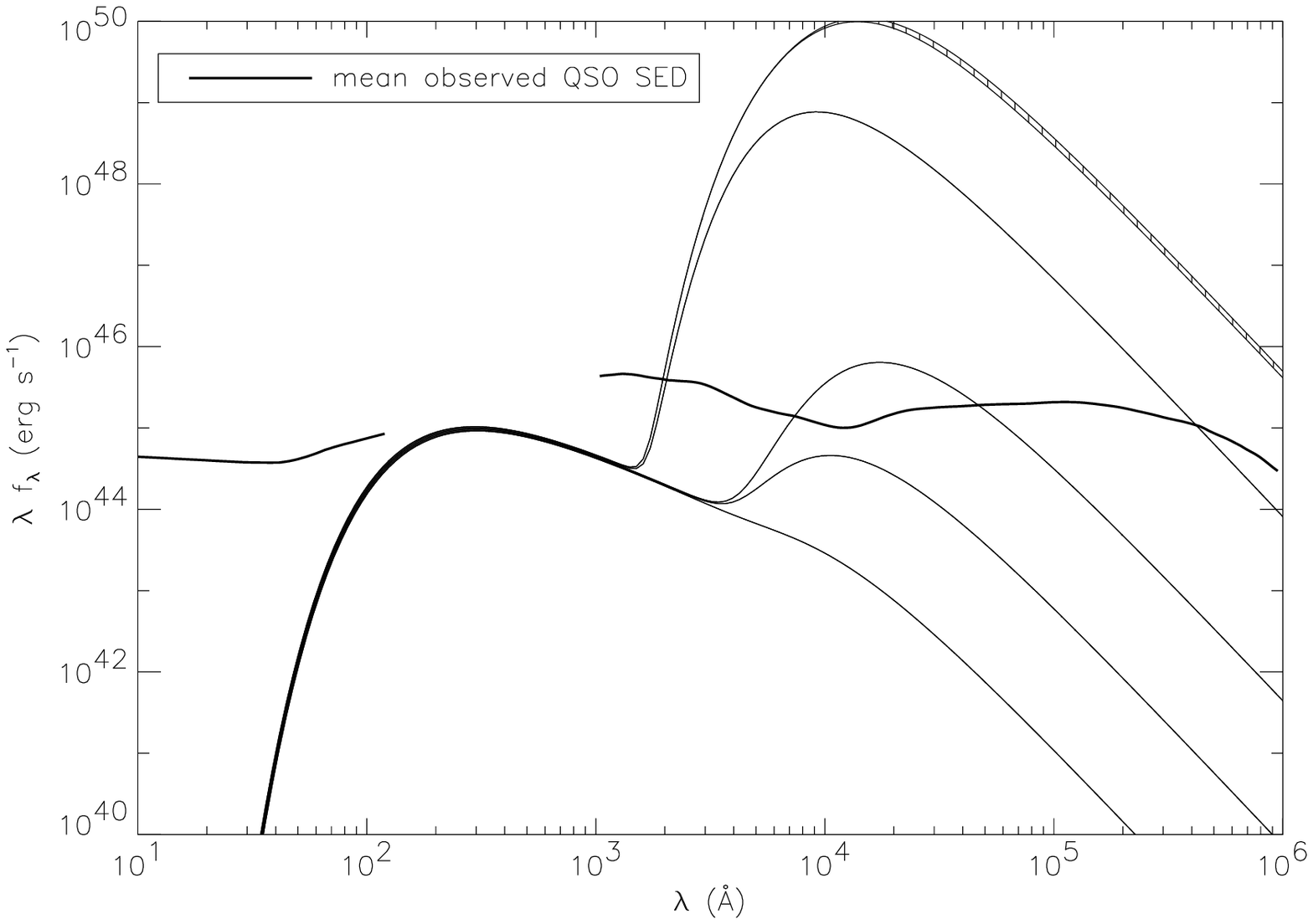}
\end{minipage}}
\caption{Calculated spectral energy distributions for our 
accretion disc models with $M_8=1$, $\lum=.5$, $\alpha=.01$.  The five
SEDs, in order of increasing total luminosity,
correspond to the five values of $\rmax/\rs = 10^3$, $10^4$, $10^5$, 
$10^6$ (diagonally-hatched), and $10^7$ (vertically-hatched).
Multiple solutions for $\Teff(r)$ are responsible for the ambiguity
of the SEDs plotted as bounded regions.}
\label{f:sed_rmax}
\end{figure}

The calculated SED of this typical model is presented in
Figure~\ref{f:sed_rmax}, 
for outer radii of $\rmax=10^3$, $10^4$, $10^5$, $10^6$, and $10^7 \rs$.
(All of our SEDs assume that the disc is viewed face-on.)
Since $\Teff$ may have multiple solutions in certain regions of the disc,
the SEDs are plotted as bounded regions with the top (bottom) 
boundary corresponding to the largest (least) $\Teff$.
Also plotted is the arbitrarily normalized mean energy distribution 
(MED) from 47 observed quasars 
\citep{Elvis_etal94}.  The authors caution that the dispersion 
about this mean is large, of order ``one decade for both the infrared and
ultraviolet components when the MED is normalized at [$\lambda=1.25 \um$].''
The $\rmax = 10^3 \rs$ case has an expected $\lambda f_\lambda \propto
\lambda^{-4/3}$ behavior since selfgravity is not an issue.
Relativistic effects \citep{laor_netzer_1989,sun_malkan_1989} only 
influence the SED at and blueward of the short wavelength peak, 
since this emission comes from the innermost parts of the disc.

\begin{figure}
\subfigure[$b=0$, $\rmax=10^5 \rs$]{
\label{f:sed_m_b0}
\begin{minipage}{.5\textwidth}
\includegraphics[width=\textwidth]{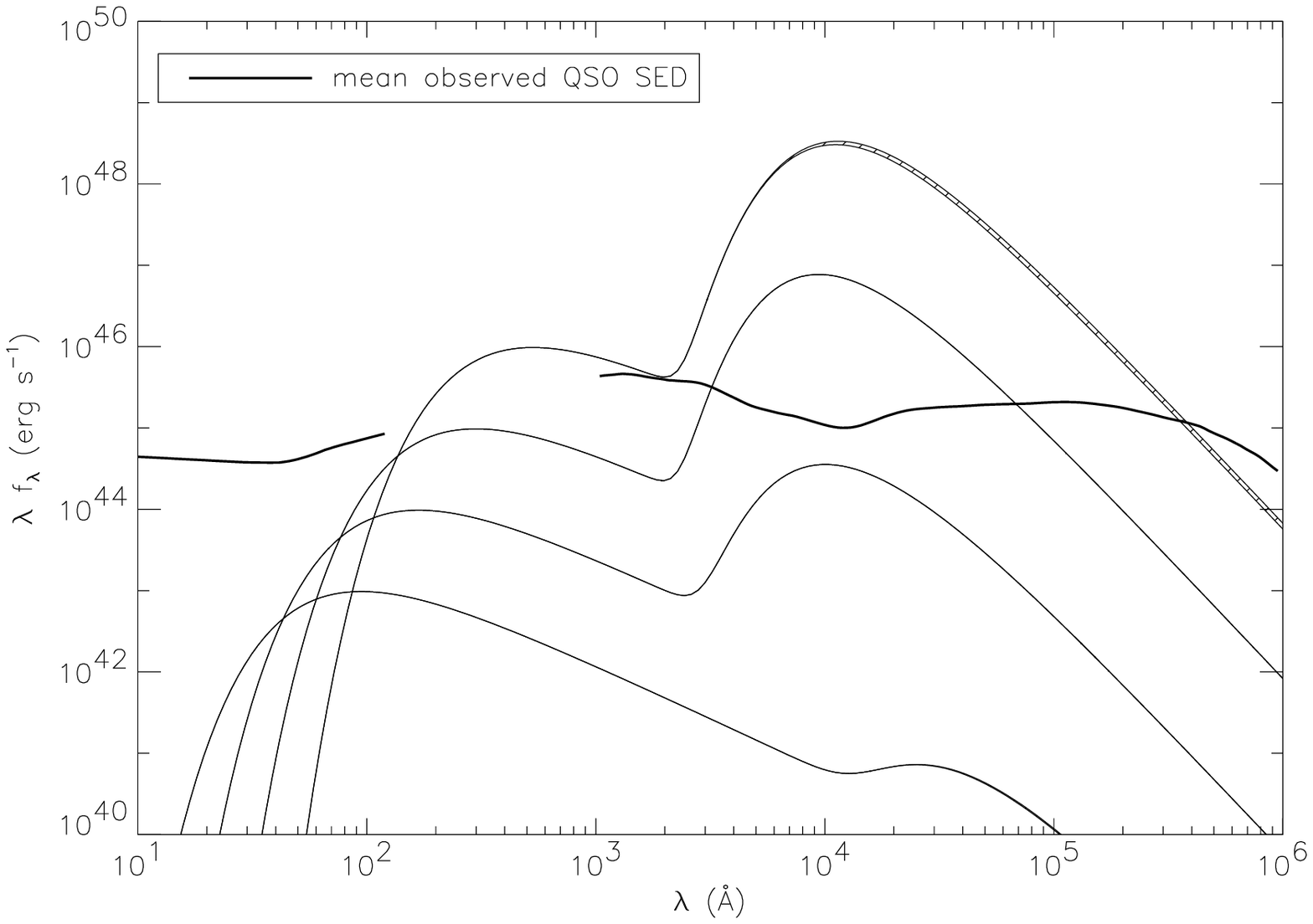}
\end{minipage}}%
\subfigure[$b=1$, $\rmax=10^5 \rs$]{
\label{f:sed_m_b1}
\begin{minipage}{.5\textwidth}
\includegraphics[width=\textwidth]{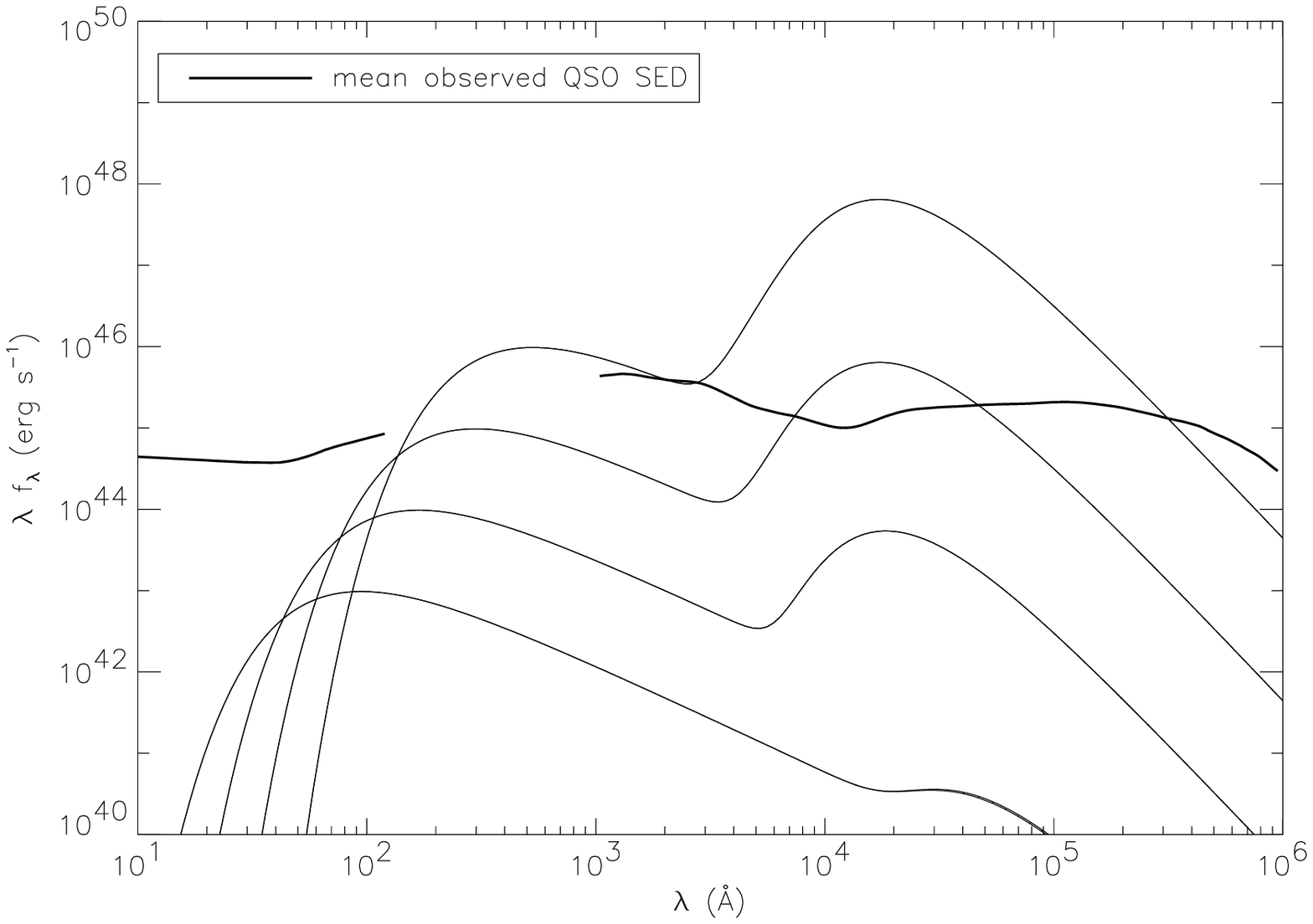}
\end{minipage}}\\
\subfigure[$b=0$, $\rmax=1\pc$]{
\label{f:sed_m_b0_rmax1pc}
\begin{minipage}{.5\textwidth}
\includegraphics[width=\textwidth]{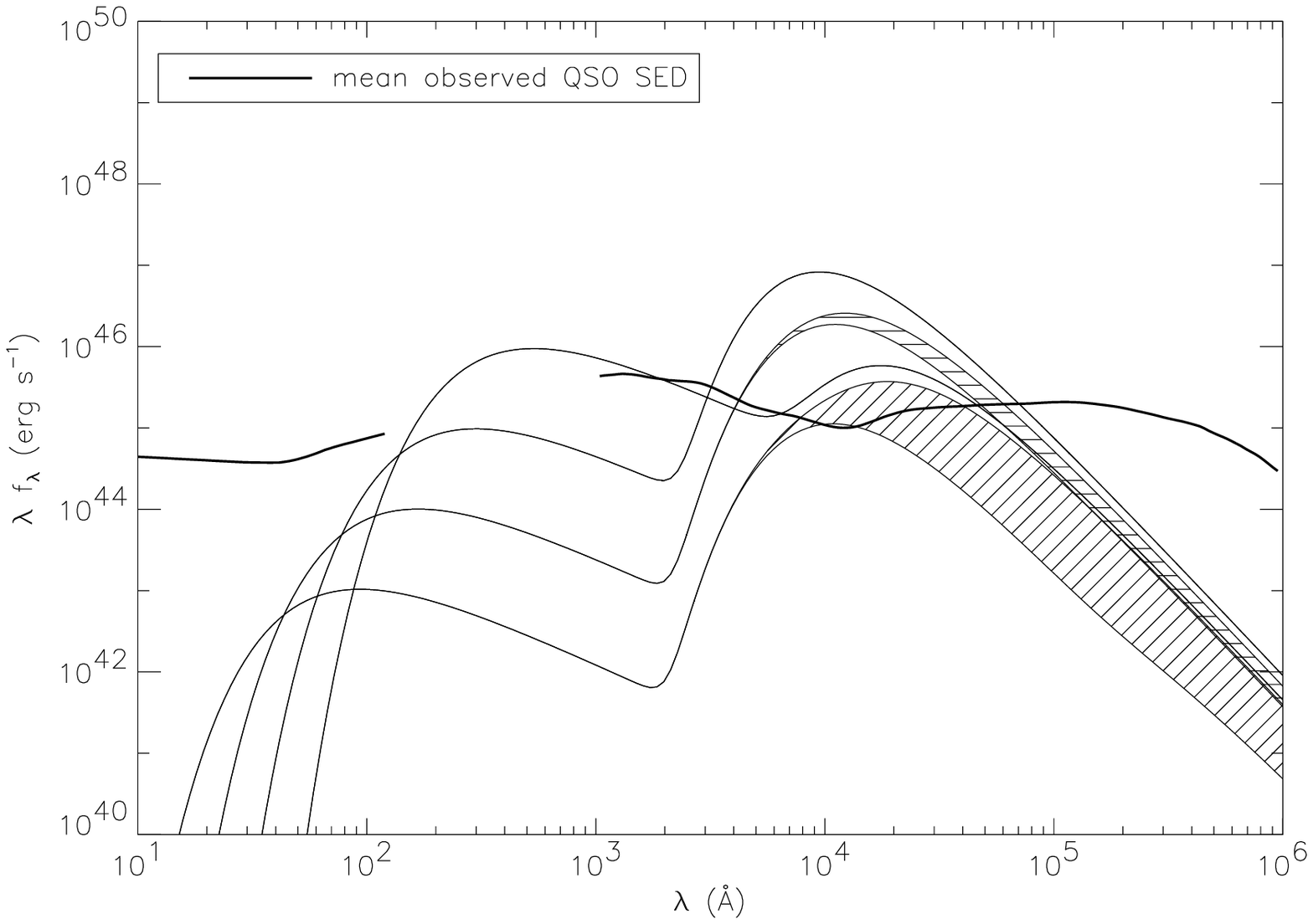}
\end{minipage}}%
\subfigure[$b=1$, $\rmax=1\pc$]{
\label{f:sed_m_b1_rmax1pc}
\begin{minipage}{.5\textwidth}
\includegraphics[width=\textwidth]{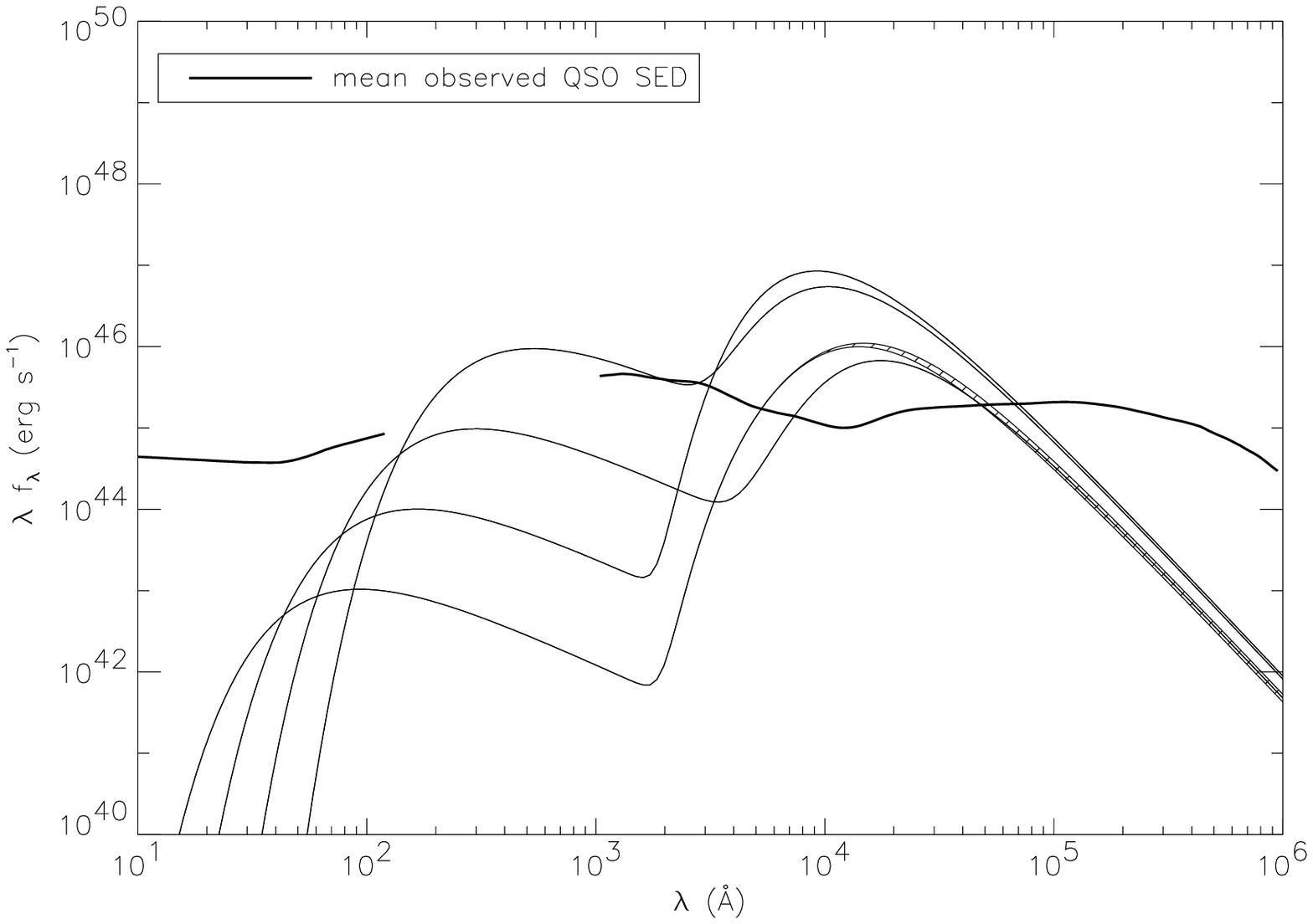}
\end{minipage}}
\caption{SEDs for the models with $\lum=.5$, $\alpha=.01$.
The four SEDs shown here, in order of increasing blue bump,
correspond to $\log{M_8} = -2$, $-1$, $0$, $1$.  The outer radius is set
to $\rmax=10^5 \rs$ in the top two panels, but it is fixed at 
$\rmax= 1\pc$ in the bottom two panels.  Compare the bottom
two panels to Figure~\protect{\ref{f:locus_m_rmax}}.}\label{f:sed_m}
\end{figure}

\begin{figure}
\includegraphics[width=\textwidth]{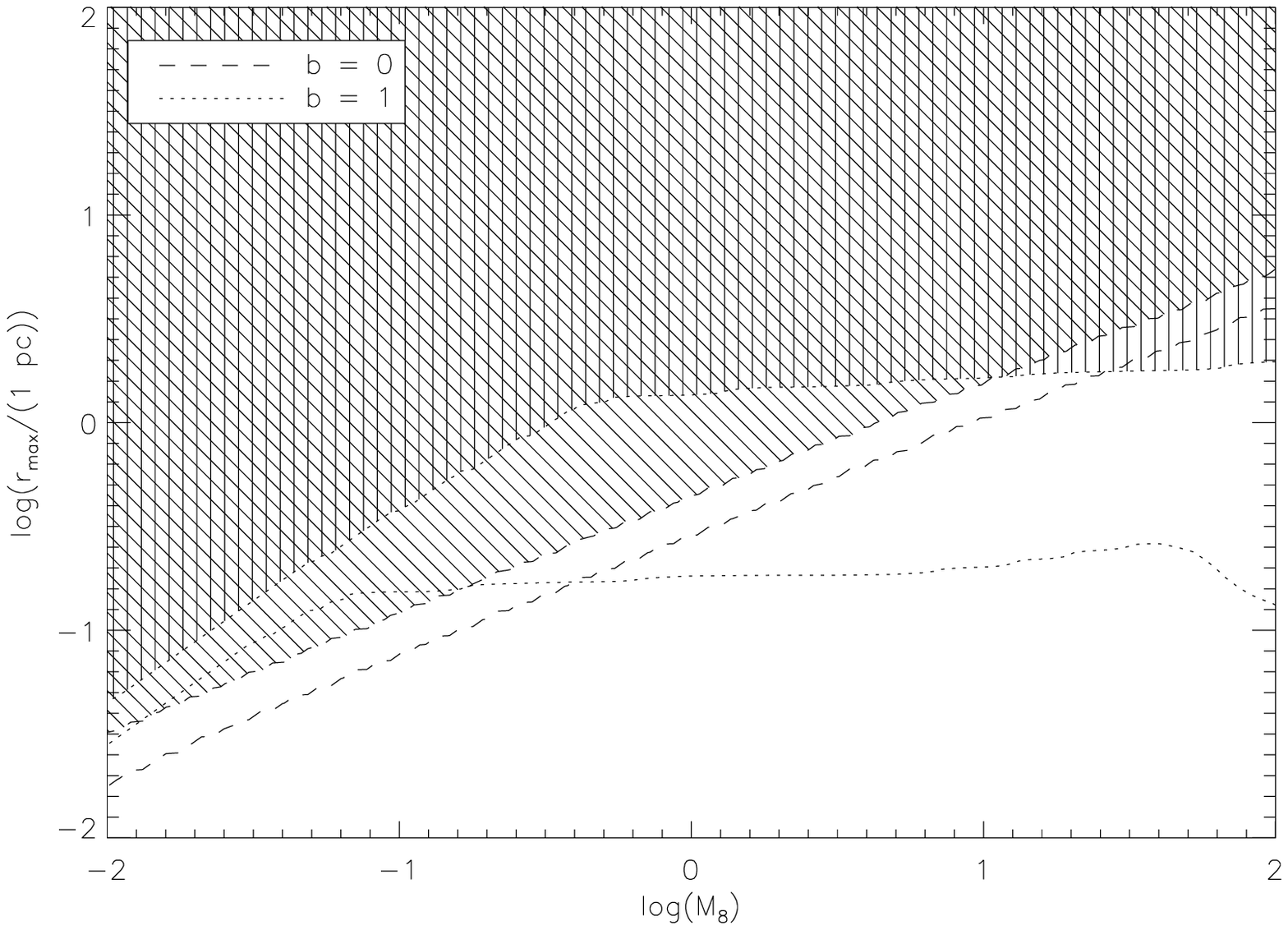}
\caption{Locus of allowable ($M$, $\rmax$), with $\lum = .5$ and 
$\alpha = .01$ fixed.  The bottom dashed (dotted) contour represents
solutions for $b=0$ ($b=1$) for which 
$(\lambda f_\lambda)_{\rm{IR\ bump}}/
	(\lambda f_\lambda)_{\rm{blue\ bump}} = 1$,
and the top contour represents solutions for which
$(\lambda f_\lambda)_{\rm{IR\ bump}}/
	(\lambda f_\lambda)_{\rm{blue\ bump}} = 10$.
Thus, the diagonally-hatched region is the observationally-constrained
approximate forbidden region for $b=0$, 
and the vertically-hatched region is the
forbidden region for $b=1$.  In both cases $\rmax$ is forbidden to
be much larger than $\sim 1 \pc$.}
\label{f:locus_m_rmax}
\end{figure}

Figure~\ref{f:sed_m} shows the dependence
of the infrared bump on the mass of
the central black hole.  (Note that the wavelength corresponding to
the blue bump has an expected $M^{1/4}$ dependence.) 
The Eddington ratio as defined above is kept constant
at $\lum = .5$: note that this parameter defines $\dot M$ rather
than the total luminosity of the disc, which includes the
auxiliary heat inputs.
In the top two panels, $\rmax$ is set to $10^5 \rs$,
and the relative prominence of the infrared bump grows with $M$
because the reduced radius at which $\Qmin=1$ decreases with $M$ (Paper I).
The bottom two panels have $\rmax$ fixed at $1 \pc$; here larger values
of $M$ tend to reduce the infrared bump.  This dependence is illustrated
in Figure~\ref{f:locus_m_rmax}.  Solutions for the disc are
considered incompatible with the observations if
$\lambda f_\lambda$ at the maximum of the infrared bump 
is more that ten times that of the blue bump.

\begin{figure}
\subfigure[b=0]{
\label{f:sed_le_b0}
\begin{minipage}{.5\textwidth}
\includegraphics[width=\textwidth]{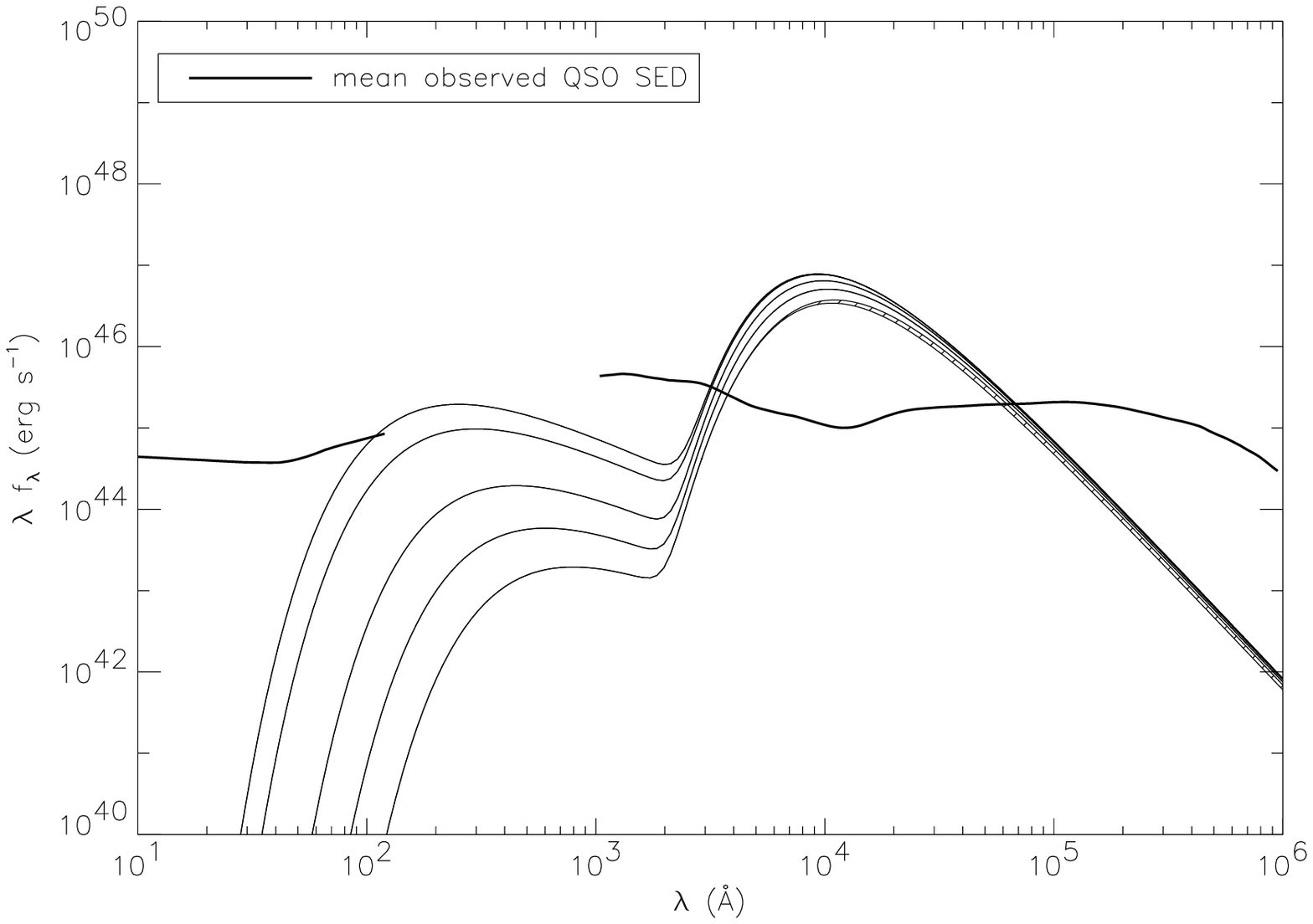}
\end{minipage}}%
\subfigure[b=1]{
\label{f:sed_le_b1}
\begin{minipage}{.5\textwidth}
\includegraphics[width=\textwidth]{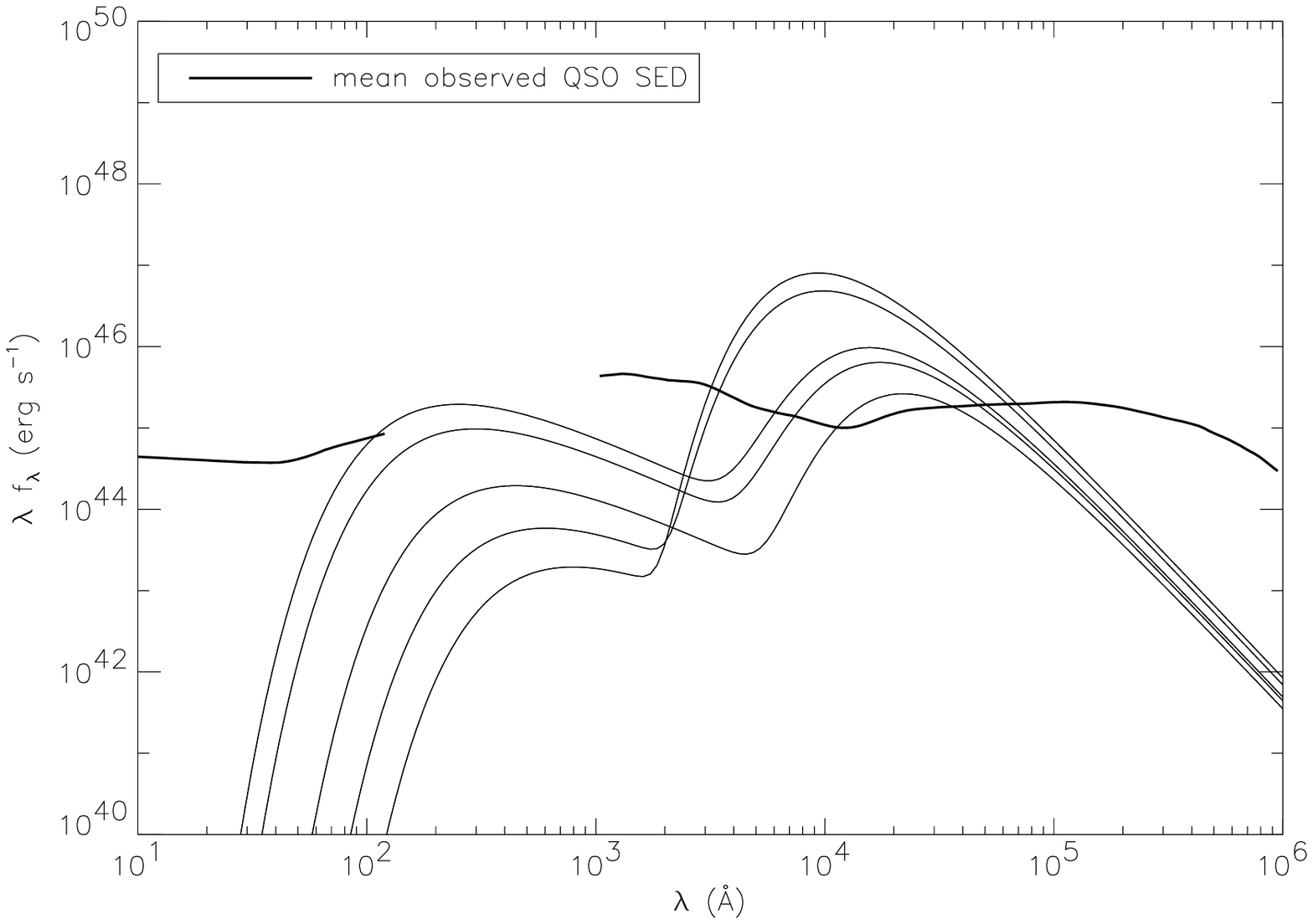}
\end{minipage}}
\caption{SEDs for the models with $M_8=1$, $\alpha=.01$, $\rmax=10^5 \rs$.
The five SEDs shown here, in order of increasing blue bump, 
correspond to $\lum = .01$, $.03$, $.1$, $.5$, $1$.}\label{f:sed_le}
\end{figure}

Figure~\ref{f:sed_le_b0} shows that while the blue bump increases
with mass accretion rate, the energy output of the infrared bump 
is hardly affected, for the $b=0$ case. 
This invariance is caused by two competing factors
as $\Mdot$ is increased: $\Teff$ is higher in the selfgravitating region, 
but because $T$ is also higher, the onset of the optically thin regime,
and thus a sharp increase in $\Teff$ as shown in 
Figure~\ref{f:radial_parms}, is delayed.
Similar considerations explain the behavior of the infrared bump in
Figure~\ref{f:sed_le_b1}, for $b=1$. For low accretion rates, the midplane 
temperature drops low enough that the disc becomes optically 
thin before the end of the disc (at $\rmax=10^5 \rs$), 
which drives up $\Teff$.

\begin{figure}
\subfigure[b=0]{
\label{f:sed_alpha_b0}
\begin{minipage}{.5\textwidth}
\includegraphics[width=\textwidth]{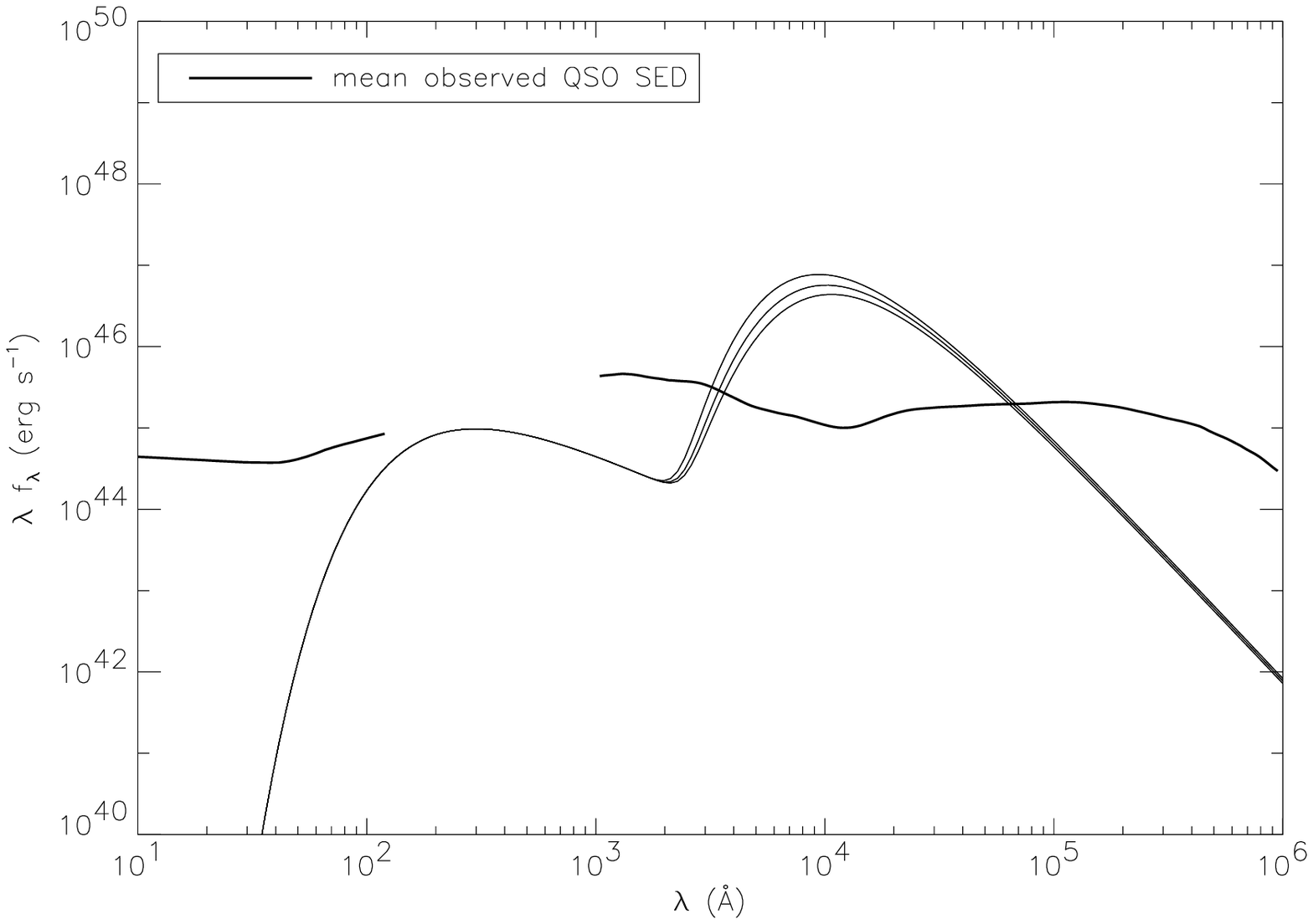}
\end{minipage}}%
\subfigure[b=1]{
\label{f:sed_alpha_b1}
\begin{minipage}{.5\textwidth}
\includegraphics[width=\textwidth]{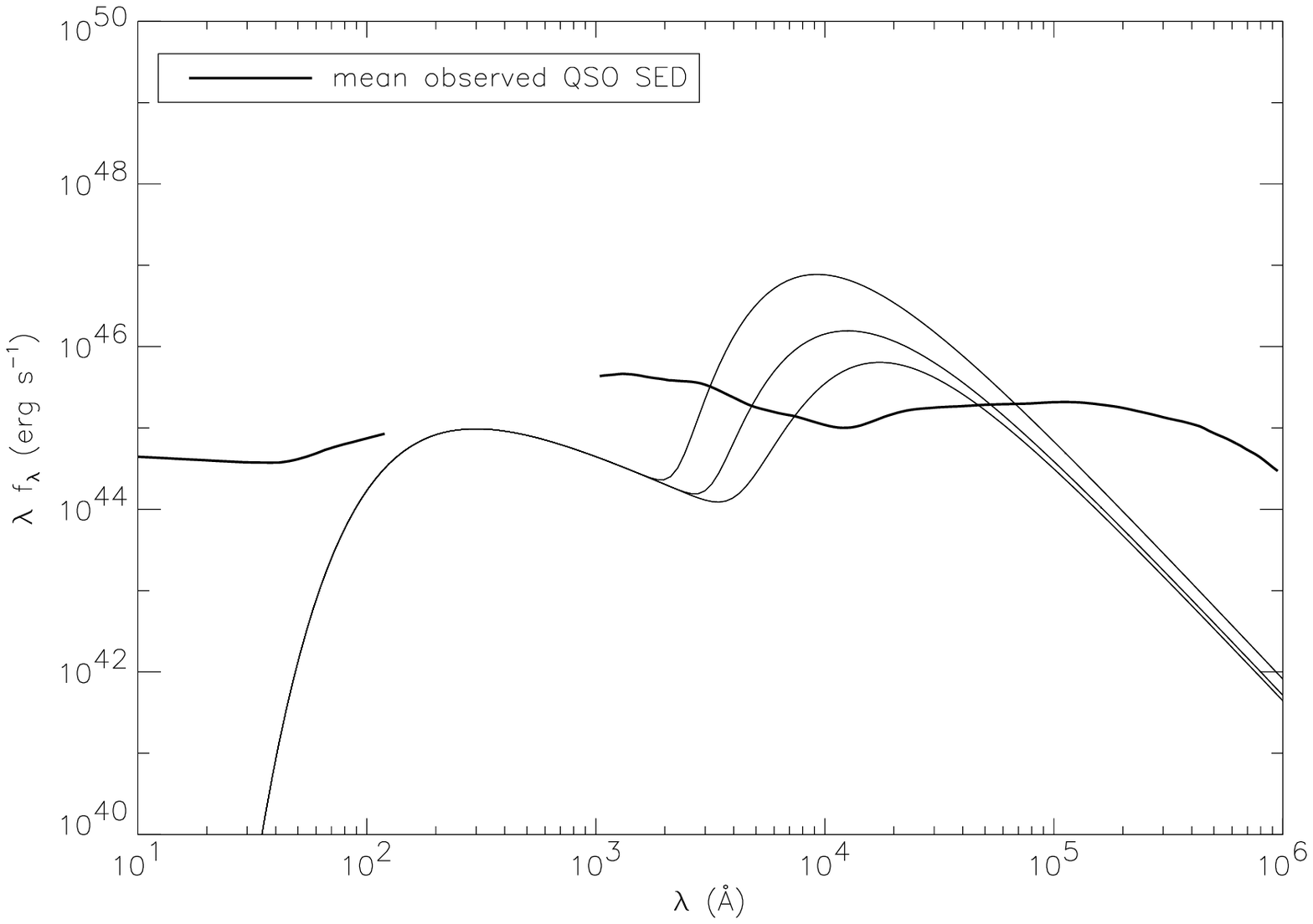}
\end{minipage}}
\caption{SEDs for the models with $M_8=1$, $\lum=.5$, $\rmax=10^5 \rs$.
The three SEDs shown here, in order of decreasing (increasing) 
infrared bump for $b=0$ ($b=1$),
correspond to $\alpha = .01$, $.1$, $.3$.}\label{f:sed_alpha}
\end{figure}

In Figure~\ref{f:sed_alpha}, the same argument explains the behavior
of the infrared bump in both the $b=0$ and $b=1$ cases. 
The onset of the optically thin selfgravitating region is 
at smaller radii with increasing $\alpha$ for both cases,
but $\Teff$ is higher in the optically thick selfgravitating region for 
$b=1$.

It is clear that additional heating in the outer parts of our accretion
discs results in a second bump in the SED.  Assuming that this 
outer-disc contribution to the SED can be no more prominent than
infrared emission in the observed MED, 
we can already place tight constraints on the outer radius of these disc
models: $\rmax \lesssim 10^4 - 10^5\rs\sim 0.1 - 1\pc$.

\section{Summary and Discussion}

We have estimated spectral energy distributions (SEDs) of bright QSOs
using standard assumptions, with one addition:  where the disc
would otherwise be gravitationally unstable, we have postulated
additional sources of heat, other than release of orbital
energy by accretion, just sufficient to maintain gravitational
stability.  These sources become necessary beyond $\sim 10^3\rs$
for typical parameters.  Assuming their energy inputs to be
locally and completely thermalized, we have calculated their
contribution to the SED and luminosity of the disc, which occurs
primarily in the red and near infrared.
The larger the disc, the more auxiliary heating is required.
For typical black-hole masses and accretion rates inferred from
the blue bump, the auxiliary inputs actually exceed the power
derived from accretion if the disc extends beyond $10^4-10^5\rs$,
or about one parsec.  This would be incompatible with the typical
SED of bright QSOs, which is approximately flat in $\lambda F_\lambda$.

Paper I placed similar limits on $r_{\rm max}$ from energetic arguments.
It was assumed that the stars or small black holes that heat
the disc also form within the disc, and that the mass in these
objects is at most comparable to that of the disc.  The present
limits do not rely on these assumptions.  That is to say, 
if the disc were heated by a much larger mass in stars, the limits
of Paper I could be relaxed, but those based on the SED would still
apply.

It has been assumed
that the accretion rate is constant with radius, so that $\dot M$ at
large radii can be derived from the luminosity in the blue bump (which
has no contribution from the auxiliary sources).  If $\dot M$ at
$r\gtrsim 1\pc$ is several orders of magnitude less than it is at
$r\lesssim 10^3\rs$, then the disc could be much more extensive than
we have supposed.  The Stefan-Boltzmann law implies
that the observed luminosity at $10\,\mu$ must in any case come from 
$r\gtrsim 2 (\lambda L_\lambda/10^{46}\mbox{ erg s}^{-1})^{1/2} \pc$,
but this does not require that the surface density at that distance
is as high as in a constant-$\dot M$ disc.
Therefore, it is useful to rephrase the limits in
terms of the initial angular momentum of the gas supplied to the disc.
At least on a time average, $\dot M$ should be constant inside the
radius corresponding to the initial angular momentum.  The
relationship is
\begin{equation}\label{J0}
J_0\approx 660 \left(r_{\pc}M_8\right)^{1/2}\pc \kms.
\end{equation}
This is quite small compared with the product of
virial velocity ($\sim 300\kms$) and scale size ($\sim 1\kpc$)
of QSO hosts, and it may be important to ask where gas with such
low angular momentum comes from.  Furthermore, the mass of
a gravitationally stable disc that obeys our constraints
on the SED is generally much less than
that of the black hole, so that the disc must be replenished
many times over to grow the black hole by accretion.

Our treatment of the disc is highly simplified and certainly crude
compared with many past efforts.  We have ignored relativistic
effects, adopted a one-zone model for the vertical structure, used
Rosseland mean opacities without distinguishing between scattering and
absorption, taken a constant molecular weight, and represented
angular-momentum transport by the usual viscous $\alpha$ prescription.
We feel that these simplifications are justified by the strong
dependence of the auxiliary inputs and the SED on the outer radius of
the disc.  A more detailed treatment of the physics that one actually
understands seems unlikely to change our conclusions concerning
$r_{\rm max}$.  Radical enhancements in transport (equivalent to
$\alpha\gg1$), or magnetic pressures $\gg \pgas$ \citep[as suggested
by][]{Pariev_etal02} could make some difference, but these are not yet
understood.

Despite what has just been said, a more elaborate treatment of
vertical structure and radiative transfer might
point to a redistribution of the auxiliary energy inputs
in wavelength, if not their total contribution to the disc luminosity.
Unfortunately, one has no reliable predictions for the vertical
distribution of purely viscous heating, much less of the auxiliary
sources postulated here, which will limit the credibility of detailed
vertical models.


Even if the actively accreting parts of QSO discs are smaller than
$0.1\pc$, it is still possible that selfgravity is important in them,
and therefore that they form stars.  Quiescent galactic nuclei with
black holes, even in early-type galaxies, often show kinematic
evidence of compact stellar discs
\citep{Gebhardt_etal00b,Bower_etal01,deZeeuw_etal02}.  Nuclear
starbursts (albeit on scales $\sim 10^2\pc$) are usually accompanied
by AGN activity \citep{Sanders99,Heckman99}.  The black hole in our
own Galaxy, though not an AGN and estimated to have a very low
accretion rate \citep{Quataert_etal99}, is surrounded by what appear
to be young high-mass stars at $r\lesssim 0.1\pc$
\citep{Krabbe_etal95}.  More theoretical attention should be paid to
star formation in these extreme environments, where the densities,
temperatures, and tidal fields are much higher than in normal giant
molecular clouds.

\bigskip
We thank Iskra Strateva and Jonathan Tan for helpful discussions.

\label{lastpage}


\begin{thebibliography}{}

\bibitem[\protect\citeauthoryear{{Alexander} \& {Ferguson}}{{Alexander} \&
  {Ferguson}}{1994}]{alexander_ferguson_1994}
{Alexander} D.~R.,  {Ferguson} J.~W.,  1994, ApJ, 437, 879

\bibitem[\protect\citeauthoryear{{Blandford} \& {Payne}}{{Blandford} \&
  {Payne}}{1982}]{Blandford_Payne82}
{Blandford} R.~D.,  {Payne} D.~G.,  1982, MNRAS, 199, 883

\bibitem[\protect\citeauthoryear{{Bower}, {Green}, {Bender}, {Gebhardt},
  {Lauer}, {Magorrian}, {Richstone}, {Danks}, {Gull}, {Hutchings}, {Joseph},
  {Kaiser}, {Weistrop}, {Woodgate}, {Nelson} \& {Malumuth}}{{Bower}
  et~al.}{2001}]{Bower_etal01}
{Bower} G.~A.,  {Green} R.~F.,  {Bender} R.,  {Gebhardt} K.,  {Lauer} T.~R.,
  {Magorrian} J.,  {Richstone} D.~O.,  {Danks} A.,  {Gull} T.,  {Hutchings} J.,
   {Joseph} C.,  {Kaiser} M.~E.,  {Weistrop} D.,  {Woodgate} B.,  {Nelson} C.,
    {Malumuth} E.~M.,  2001, ApJ, 550, 75

\bibitem[\protect\citeauthoryear{{Chokshi} \& {Turner}}{{Chokshi} \&
  {Turner}}{1992}]{Chokshi_Turner92}
{Chokshi} A.,  {Turner} E.~L.,  1992, MNRAS, 259, 421

\bibitem[\protect\citeauthoryear{{Collin} \& {Zahn}}{{Collin} \&
  {Zahn}}{1999a}]{Collin_Zahn99a}
{Collin} S.,  {Zahn} J.,  1999a, A\&A, 344, 433

\bibitem[\protect\citeauthoryear{{Collin} \& {Zahn}}{{Collin} \&
  {Zahn}}{1999b}]{Collin_Zahn99b}
{Collin} S.,  {Zahn} J.,  1999b, Ap\&SS, 265, 501

\bibitem[\protect\citeauthoryear{{de Zeeuw}, {Bureau}, {Emsellem}, {Bacon},
  {Marcella Carollo}, {Copin}, {Davies}, {Kuntschner}, {Miller}, {Monnet},
  {Peletier} \& {Verolme}}{{de Zeeuw} et~al.}{2002}]{deZeeuw_etal02}
{de Zeeuw} P.~T.,  {Bureau} M.,  {Emsellem} E.,  {Bacon} R.,  {Marcella
  Carollo} C.,  {Copin} Y.,  {Davies} R.~L.,  {Kuntschner} H.,  {Miller} B.~W.,
   {Monnet} G.,  {Peletier} R.~F.,    {Verolme} E.~K.,  2002, MNRAS, 329, 513

\bibitem[\protect\citeauthoryear{{Elvis}, {Wilkes}, {McDowell}, {Green},
  {Bechtold}, {Willner}, {Oey}, {Polomski} \& {Cutri}}{{Elvis}
  et~al.}{1994}]{Elvis_etal94}
{Elvis} M.,  {Wilkes} B.~J.,  {McDowell} J.~C.,  {Green} R.~F.,  {Bechtold} J.,
   {Willner} S.~P.,  {Oey} M.~S.,  {Polomski} E.,    {Cutri} R.,  1994, ApJS,
  95, 1

\bibitem[\protect\citeauthoryear{{Eracleous} \& {Halpern}}{{Eracleous} \&
  {Halpern}}{1994}]{Eracleous_Halpern94}
{Eracleous} M.,  {Halpern} J.~P.,  1994, ApJS, 90, 1

\bibitem[\protect\citeauthoryear{{Gebhardt}, {Richstone}, {Kormendy}, {Lauer},
  {Ajhar}, {Bender}, {Dressler}, {Faber}, {Grillmair}, {Magorrian} \&
  {Tremaine}}{{Gebhardt} et~al.}{2000}]{Gebhardt_etal00b}
{Gebhardt} K.,  {Richstone} D.,  {Kormendy} J.,  {Lauer} T.~R.,  {Ajhar} E.~A.,
   {Bender} R.,  {Dressler} A.,  {Faber} S.~M.,  {Grillmair} C.,  {Magorrian}
  J.,    {Tremaine} S.,  2000, AJ, 119, 1157

\bibitem[\protect\citeauthoryear{Goodman}{Goodman}{2002}]{Goodman02}
Goodman J.,  2002, astro-ph/0201001

\bibitem[\protect\citeauthoryear{{Heckman}}{{Heckman}}{1999}]{Heckman99}
{Heckman} T.~M.,  1999, in IAU Symp. 193: Wolf-Rayet Phenomena in Massive Stars
  and Starburst Galaxies Vol.~193, {The energetic role of massive stars in the
  AGN phenomenon}.
pp 703--715

\bibitem[\protect\citeauthoryear{{Iglesias} \& {Rogers}}{{Iglesias} \&
  {Rogers}}{1996}]{iglesias_rogers_1996}
{Iglesias} C.~A.,  {Rogers} F.~J.,  1996, ApJ, 464, 943

\bibitem[\protect\citeauthoryear{{Krabbe}, {Genzel}, {Eckart}, {Najarro},
  {Lutz}, {Cameron}, {Kroker}, {Tacconi-Garman}, {Thatte}, {Weitzel},
  {Drapatz}, {Geballe}, {Sternberg} \& {Kudritzki}}{{Krabbe}
  et~al.}{1995}]{Krabbe_etal95}
{Krabbe} A.,  {Genzel} R.,  {Eckart} A.,  {Najarro} F.,  {Lutz} D.,  {Cameron}
  M.,  {Kroker} H.,  {Tacconi-Garman} L.~E.,  {Thatte} N.,  {Weitzel} L.,
  {Drapatz} S.,  {Geballe} T.,  {Sternberg} A.,    {Kudritzki} R.,  1995, ApJ,
  447, L95

\bibitem[\protect\citeauthoryear{{Laor} \& {Netzer}}{{Laor} \&
  {Netzer}}{1989}]{laor_netzer_1989}
{Laor} A.,  {Netzer} H.,  1989, MNRAS, 238, 897

\bibitem[\protect\citeauthoryear{Pariev, Blackman \& Boldyrev}{Pariev
  et~al.}{2002}]{Pariev_etal02}
Pariev V.~I.,  Blackman E.~G.,    Boldyrev S.~A.,  2002, astro-ph/0208400

\bibitem[\protect\citeauthoryear{{Pringle}}{{Pringle}}{1981}]{Pringle81}
{Pringle} J.~E.,  1981, ARAA, 19, 137

\bibitem[\protect\citeauthoryear{{Quataert}, {Narayan} \& {Reid}}{{Quataert}
  et~al.}{1999}]{Quataert_etal99}
{Quataert} E.,  {Narayan} R.,    {Reid} M.~J.,  1999, ApJ, 517, L101

\bibitem[\protect\citeauthoryear{{Sanders}}{{Sanders}}{1999}]{Sanders99}
{Sanders} D.~B.,  1999, Ap\&SS, 266, 331

\bibitem[\protect\citeauthoryear{{Sanders}, {Phinney}, {Neugebauer}, {Soifer}
  \& {Matthews}}{{Sanders} et~al.}{1989}]{Sanders_etal89}
{Sanders} D.~B.,  {Phinney} E.~S.,  {Neugebauer} G.,  {Soifer} B.~T.,
  {Matthews} K.,  1989, ApJ, 347, 29

\bibitem[\protect\citeauthoryear{{Shapiro}, {Lightman} \& {Eardley}}{{Shapiro}
  et~al.}{1976}]{Shapiro_Lightman_Eardley76}
{Shapiro} S.~L.,  {Lightman} A.~P.,    {Eardley} D.~M.,  1976, ApJ, 204, 187

\bibitem[\protect\citeauthoryear{{Shlosman} \& {Begelman}}{{Shlosman} \&
  {Begelman}}{1987}]{Shlosman_Begelman87}
{Shlosman} I.,  {Begelman} M.~C.,  1987, Nature, 329, 810

\bibitem[\protect\citeauthoryear{{Shlosman} \& {Begelman}}{{Shlosman} \&
  {Begelman}}{1989}]{Shlosman_Begelman89}
{Shlosman} I.,  {Begelman} M.~C.,  1989, ApJ, 341, 685

\bibitem[\protect\citeauthoryear{{Siemiginowska}, {Czerny} \&
  {Kostyunin}}{{Siemiginowska} et~al.}{1996}]{siemiginowska_etal_1996}
{Siemiginowska} A.,  {Czerny} B.,    {Kostyunin} V.,  1996, ApJ, 458, 491

\bibitem[\protect\citeauthoryear{{Soltan}}{{Soltan}}{1982}]{Soltan82}
{Soltan} A.,  1982, MNRAS, 200, 115

\bibitem[\protect\citeauthoryear{{Sun} \& {Malkan}}{{Sun} \&
  {Malkan}}{1989}]{sun_malkan_1989}
{Sun} W.,  {Malkan} M.~A.,  1989, ApJ, 346, 68

\bibitem[\protect\citeauthoryear{{Yu} \& {Tremaine}}{{Yu} \&
  {Tremaine}}{2002}]{Yu_Tremaine02}
{Yu} Q.,  {Tremaine} S.,  2002, MNRAS, 335, 965

\end{thebibliography}
\end{document}